\documentclass[aps,prl,twocolumn,superscriptaddress,floatfix,nofootinbib,showpacs,amsmath,amssymb,altaffilletter]{revtex4-1}
\usepackage[colorlinks=true,pdfstartview=FitV,linkcolor=blue,citecolor=blue, urlcolor=blue]{hyperref}
\usepackage{rotate}
\usepackage{rotating}
\usepackage{graphicx}
\usepackage{bm}
\usepackage{wrapfig}
\usepackage{paralist}
\usepackage{wrapfig}
\usepackage{subfigure}
\usepackage{caption2}
\usepackage{longtable}
\usepackage{color}
\usepackage{appendix}
\usepackage{vruler}
\usepackage{hyperref}
\usepackage{setspace}
\usepackage{ifthen}
\usepackage{array}
\usepackage{etoolbox}
\usepackage[sc]{mathpazo}
\newcommand{\myref}[2]{\href{http://dx.doi.org/#2}{#1}}
\newcommand{\mref}[1]{\href{http://#1}{#1}}

\newcommand{\bg}{\mbox{$\beta/\gamma$}}

\newcommand{\sodium}{\mbox{$^{22}$Na}}

\newcommand{\rbthree}{\mbox{$^{83}$Rb}}
\newcommand{\krthree}{\mbox{$^{83m}$Kr}}

\newcommand{\kevr}{\mbox{keV}}

\newcommand{\fno}{\mbox{\tt f$_{90}$}}
\newcommand{\pe}{\mbox{PE}}

\newcommand{\dst}{DarkSide-10}

\newcommand{\lartpc}{LAr-TPC}

\newcommand{\tpctof}{TPCtof}
\newcommand{\ntof}{Ntof}
\newcommand{\npsd}{Npsd}
\begin{document}

\title{Observation of the Dependence of Scintillation from Nuclear Recoils in Liquid Argon on Drift Field}

\newcommand{\FNAL}{Fermi National Accelerator Laboratory, Batavia, IL 60510, USA}
\newcommand{\Houston}{Department of Physics, University of Houston, Houston, TX 77204, USA}
\newcommand{\LNGS}{INFN Laboratori Nazionali del Gran Sasso, Assergi 67010, Italy}
\newcommand{\Napoli}{Physics Department, Universit\`a degli Studi Federico II and INFN, Napoli 80126, Italy}
\newcommand{\NotreDame}{Physics Department, University of Notre Dame, Notre Dame, IN 46556, USA}
\newcommand{\Princeton}{Physics Department, Princeton University, Princeton, NJ 08544, USA}
\newcommand{\Temple}{Physics Department, Temple University, Philadelphia, PA 19122, USA}
\newcommand{\UCL}{Department of Physics and Astronomy, University College London, London WC1E 6BT, United Kingdom}
\newcommand{\UCLA}{Physics and Astronomy Department, University of California, Los Angeles, CA 90095, USA}
\newcommand{\UMass}{Physics Department, University of Massachusetts, Amherst, MA 01003, USA}
\newcommand{\KICPChicago}{Kavli Institute for Cosmological Physics, University of Chicago, Chicago, IL 60637, USA}

%author list as of 117:35 6/12/13
\author{T.~Alexander}\affiliation{Physics Department, University of Massachusetts, Amherst, MA 01003, USA}\affiliation{Fermi National Accelerator Laboratory, Batavia, IL 60510, USA}
\author{H.~O.~Back}\affiliation{Physics Department, Princeton University, Princeton, NJ 08544, USA}
\author{H.~Cao}\affiliation{Physics Department, Princeton University, Princeton, NJ 08544, USA}
\author{A.~G.~Cocco}\affiliation{Physics Department, Universit\`a degli Studi Federico II and INFN, Napoli 80126, Italy}
\author{F.~DeJongh}\affiliation{Fermi National Accelerator Laboratory, Batavia, IL 60510, USA}
\author{G.~Fiorillo}\affiliation{Physics Department, Universit\`a degli Studi Federico II and INFN, Napoli 80126, Italy}
\author{C.~Galbiati}\affiliation{Physics Department, Princeton University, Princeton, NJ 08544, USA}
\author{L.~Grandi}\affiliation{Kavli Institute for Cosmological Physics, University of Chicago, Chicago, IL 60637, USA}\affiliation{Physics Department, Princeton University, Princeton, NJ 08544, USA}
\author{C.~Kendziora}\affiliation{Fermi National Accelerator Laboratory, Batavia, IL 60510, USA}
\author{W.~H.~Lippincott}\affiliation{Fermi National Accelerator Laboratory, Batavia, IL 60510, USA}
\author{B.~Loer}\affiliation{Fermi National Accelerator Laboratory, Batavia, IL 60510, USA}
\author{C.~Love}\affiliation{Physics Department, Temple University, Philadelphia, PA 19122, USA}
\author{L.~Manenti}\affiliation{Department of Physics and Astronomy, University College London, London WC1E 6BT, United Kingdom}
\author{C.~J.~Martoff}\affiliation{Physics Department, Temple University, Philadelphia, PA 19122, USA}
\author{Y.~Meng}\affiliation{Physics and Astronomy Department, University of California, Los Angeles, CA 90095, USA}
\author{D.~Montanari}\affiliation{Fermi National Accelerator Laboratory, Batavia, IL 60510, USA}
\author{P.~Mosteiro}\affiliation{Physics Department, Princeton University, Princeton, NJ 08544, USA}
\author{D.~Olvitt}\affiliation{Physics Department, Temple University, Philadelphia, PA 19122, USA}
\author{S.~Pordes}\affiliation{Fermi National Accelerator Laboratory, Batavia, IL 60510, USA}
\author{H.~Qian}\affiliation{Physics Department, Princeton University, Princeton, NJ 08544, USA}
\author{B.~Rossi}\affiliation{Physics Department, Universit\`a degli Studi Federico II and INFN, Napoli 80126, Italy}\affiliation{Physics Department, Princeton University, Princeton, NJ 08544, USA}
\author{R.~Saldanha}\affiliation{Physics Department, Princeton University, Princeton, NJ 08544, USA}\affiliation{INFN Laboratori Nazionali del Gran Sasso, Assergi 67010, Italy}
\author{W.~Tan}\affiliation{Physics Department, University of Notre Dame, Notre Dame, IN 46556, USA}
\author{J.~Tatarowicz}\affiliation{Physics Department, Temple University, Philadelphia, PA 19122, USA}
\author{S.~Walker}\affiliation{Physics Department, Temple University, Philadelphia, PA 19122, USA}
\author{H.~Wang}\affiliation{Physics and Astronomy Department, University of California, Los Angeles, CA 90095, USA}
\author{A.~W.~Watson}\affiliation{Physics Department, Temple University, Philadelphia, PA 19122, USA}
\author{S.~Westerdale}\affiliation{Physics Department, Princeton University, Princeton, NJ 08544, USA}
\author{J.~Yoo}\affiliation{Fermi National Accelerator Laboratory, Batavia, IL 60510, USA}

\collaboration{The SCENE Collaboration}

\keywords{Dark Matter; Noble Liquid TPC; Liquid Argon TPC, Scintillation}
\pacs{29.40.Cs, 32.10.Hq, 34.90.+q, 51.50.+v, 52.20.Hv}

%%%%%%%%%
\begin{abstract}

We have exposed a dual-phase Liquid Argon Time Projection Chamber (\lartpc) to a low energy pulsed narrowband neutron beam, produced at the Notre Dame Institute for Structure and Nuclear Astrophysics, to study the scintillation light yield of recoiling nuclei.  Liquid scintillation counters were arranged to detect and identify neutrons scattered in the \lartpc\ and to select the energy of the recoiling nuclei. 

We report the observation of a significant dependence (up to 32$\%$) on drift field of liquid argon scintillation from nuclear recoils of energies between 10.8\,\kevr and 49.9\,\kevr.  The field dependence is stronger at lower energies. The first measurement of such an effect in liquid argon, this observation is important because, to date, estimates of the sensitivity of LAr-TPC dark matter searches are based on the assumption that electric field has only a small effect on the light yield from nuclear recoils. 
\end{abstract}

\maketitle

Noble liquid TPCs are in widespread use to search for WIMP dark matter by detecting low energy nuclear recoils that would be produced by WIMP interactions~\cite{xenon,lux,ardm,warp,darkside,pandax}.  A precise understanding of the scintillation light yield from nuclear recoils and its dependence on energy and drift field are key to the interpretation of results from present experiments and to estimates of sensitivity of future detectors.

%Two methods have been used to measure the light yield from nuclear recoils. In the first, monoenergetic neutron sources scatter into an angle selected by an external neutron detector, producing a specific recoil energy in the target fluid. In the second, data from broad spectrum neutron sources such as AmBe are fit to simulations.  
Measurements of the nuclear recoil light yield in liquid argon (LAr) have been performed in absence of a drift field with monoenergetic neutrons and reported in~\cite{gastler,regenfus}.  For liquid xenon (LXe), several measurements have been performed with and without an applied electric field using both monoenergetic and broad spectrum neutron sources~\cite{arneodo,gokalp,akimov,aprile2005,shutt,chepel,aprile2009,sorensen,lebedenko,manzur,plante,horn}. These measurements in LXe report that the applied electric field has a $<10\%$ effect on the nuclear recoil light yield that is independent of energy, although the uncertainty increases as the recoil energy decreases. 

%Two such experiments were able to apply an electric field and observed a $<$10$\%$ decrease in the light yield due to an applied drift field for 56\,\kevr\,recoils ~\cite{aprile2005} although with large uncertainty at low energies~\cite{manzur}. Based on these results, published dark matter limits from the XENON100 collaboration assume an energy-independent $5\%$ decrease of the light yield due to a 530 V/cm applied field, and this assumption is supported by a recent broad spectrum measurement made in the presence of an electric field~\cite{Aprile2013}.

We have used a monoenergetic neutron beam to observe the light produced by nuclear recoils between 10.8 and 49.9\,\kevr\ in a LAr time projection chamber (LAr-TPC) with and without an applied electric field. We report a strong dependence on drift field of the scintillation yield, a dependence that increases with decreasing energy.

The experiment was performed at the University of Notre Dame Institute for Structure and Nuclear Astrophysics.  Protons from the Tandem accelerator~\cite{tandem} struck a 0.20\,mg/cm$^2$ thick LiF target deposited on a 1-mm-thick aluminum backing, generating a neutron beam through the reaction $^7$Li(p,n)$^7$Be. For this study, the proton beam energy was 2.376 or 2.930\,MeV, depending on the desired nuclear recoil energy.  The proton beam was bunched and chopped to provide pulses 1\,ns wide, separated by 101.5\,ns, with an average of $6.3\times10^4$ protons per pulse. The accelerator pulse selector was set to allow one of every two proton bunches to strike the LiF target, giving one neutron beam pulse every 203.0\,ns.  

%The average energy of neutrons hitting the \lartpc\ was 604\,keV or 1.168\,MeV, with a spread at zero degrees of $\sim$2\,keV FWHM~\cite{cao-mosteiro}. 

The design of the LAr-TPC closely follows that used in \dst~\cite{darkside}.  The active volume is contained within a 68.6\,mm diameter, 76.2\,mm tall, right circular polytetrafluoroethylene (PTFE) cylinder lined with 3M Vikuiti enhanced specular reflector ~\cite{vikuiti} and capped by fused silica windows. The LAr is viewed through the windows by two 3'' Hamamatsu R11065 PhotoMultiplier Tubes (PMTs)~\cite{hamamatsu}. The windows are coated with the transparent conductive material indium tin oxide (ITO), allowing for the application of electric field, and copper field rings embedded in the PTFE cylinder maintained field uniformity.  All internal surfaces of the detector were coated with the wavelength shifter TetraPhenylButadiene (TPB). The LAr-TPC was located 73.1 cm from the LiF target, and the average number of neutrons passing through the TPC per pulse was $2.9\times10^{-4}$.
%A diaphragm pump circulated boil-off argon gas through a purifier before the gas was recondensed into the active volume by a cryocooler.   

\begin{figure}[t!]
\includegraphics[width=\columnwidth]{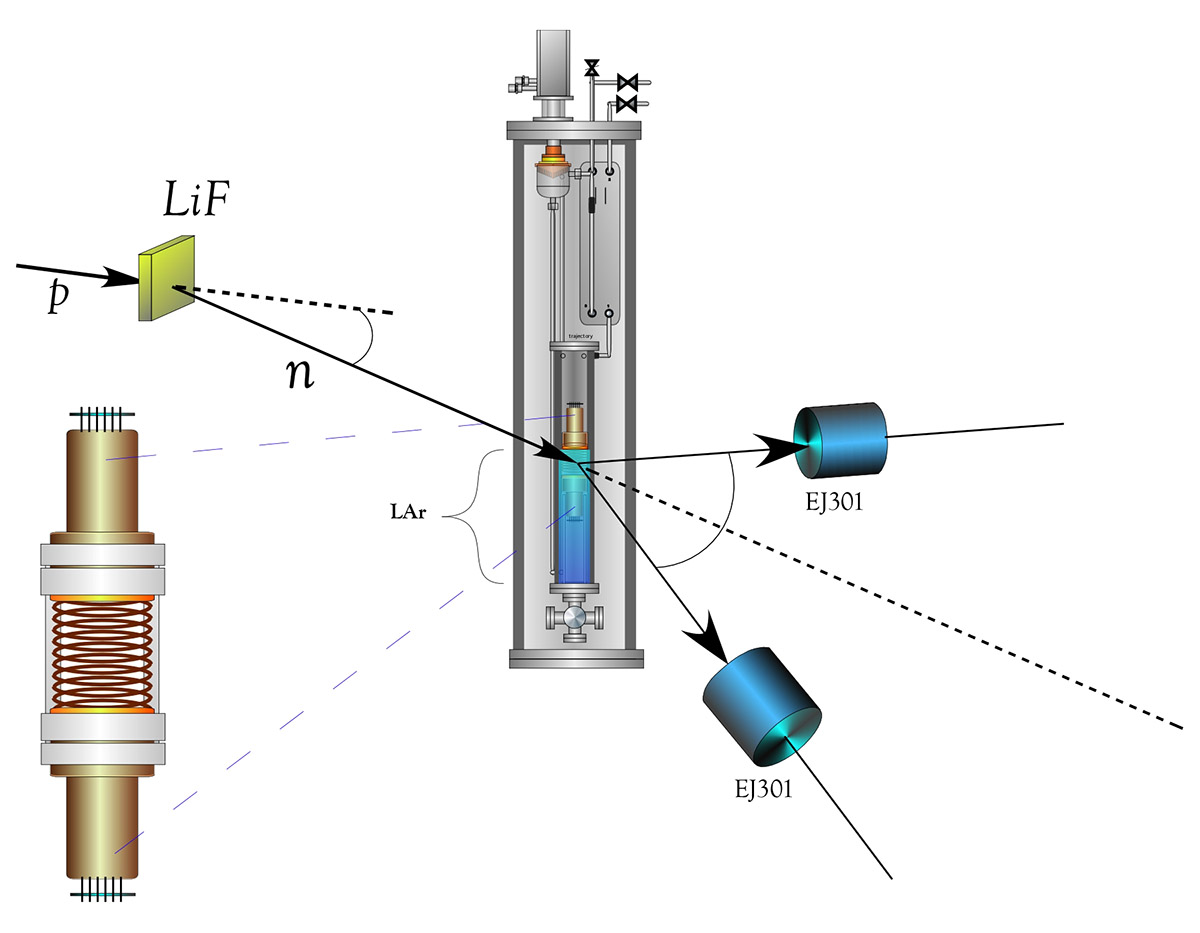}
\caption{\label{fig:schematic}A schematic of the experiment setup. The zoomed-in view of the TPC shows the PMTs, field shaping rings and PTFE support structure. It does not include the inner reflector.}
\end{figure}

Scattered neutrons were detected in two 12.7$\times$12.7\,cm cylindrical liquid scintillator counters~\cite{eljen}. These detectors were placed 71\,cm from the LAr target at a variety of angles to the beam direction, thus defining the nuclear recoil energy in the TPC. Figure~\ref{fig:schematic} shows a schematic of the geometry along with a zoomed-in view of the TPC, and  Table~\ref{table:neutrons} lists the configurations of beam energy and detector location and the corresponding mean nuclear recoil energy in the LAr-TPC.  The scintillators provide timing information and pulse shape discrimination, both of which suppress background from $\gamma$-ray interactions.  Cylinders of polyethylene (22$\times$22\,cm) shielded the neutron detectors from direct view of the LiF target for all but the 49.9\,\kevr\ data.

\begin{table}
 \begin{center} 
 \begin{tabular}{|c|c|c|c|} \hline
   Proton & Neutron & Scattering & Nuclear recoil \\
 energy & energy & angle & energy  \\
 (MeV) & (MeV) & ($^\circ$) & (keVr) \\ \hline
    2.376 & 0.604 & 49.9 & 10.8 \\
  2.930 & 1.168 & 42.2 & 15.2 \\
  2.930 & 1.168 & 49.9 & 20.8 \\
  2.930 & 1.168 & 59.9 & 29.0 \\
  2.930 & 1.168 & 82.2 & 49.9\\ \hline
\end{tabular}
 \caption{Run parameters for the data presented in this letter.}
 \label{table:neutrons}
 \end{center}
\end{table}

\begin{figure}[t!]
\includegraphics[width=\columnwidth]{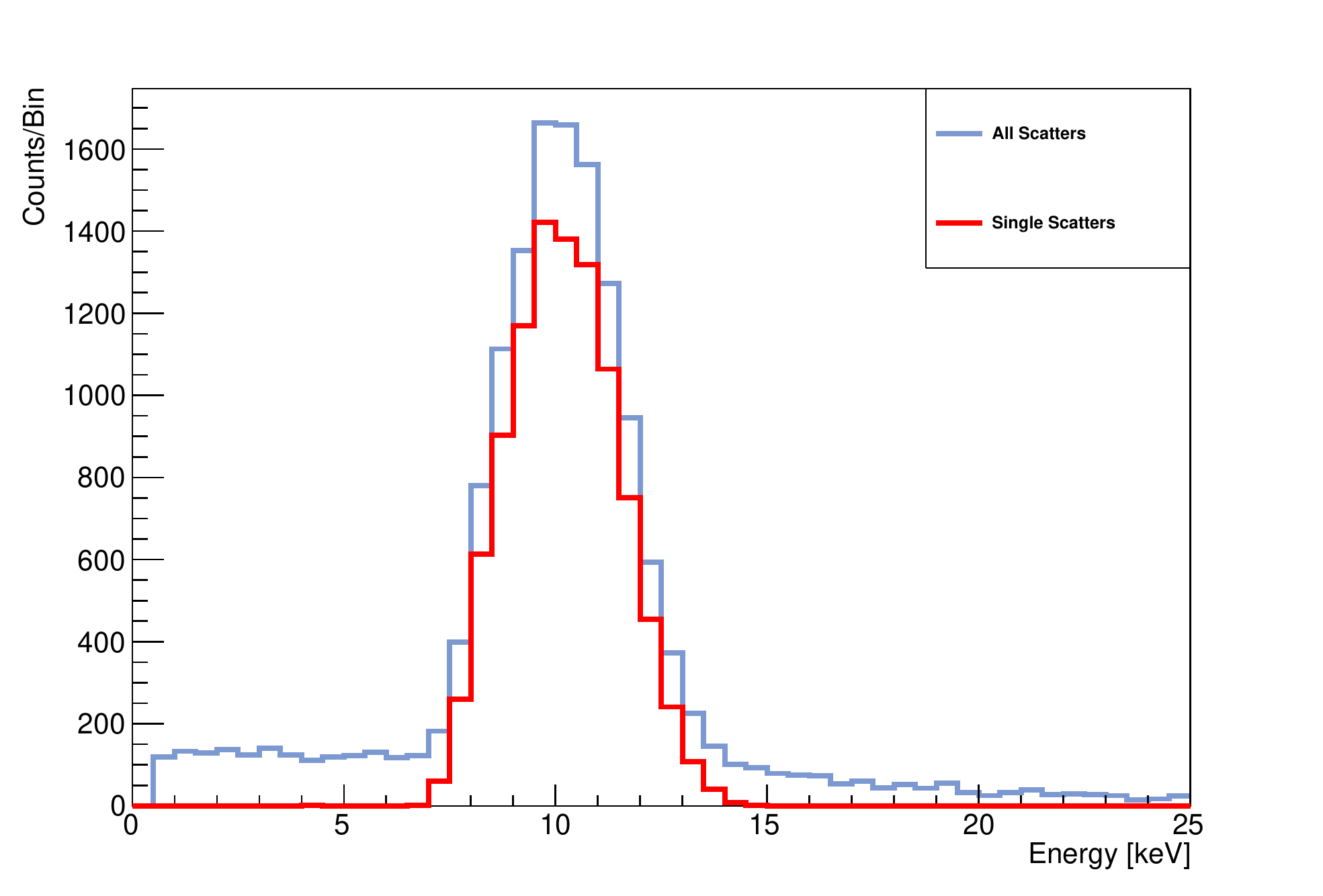}
\caption{\label{fig:mc}GEANT4-based simulation of the energy deposition in the SCENE detector at the 10.8\,\kevr\,setting.  Blue: all scatters in the LAr-TPC producing coincidences in either neutron detector and surviving the timing cuts discussed in the text.  Red: single elastic scatters surviving the same cuts.}
\end{figure}

Our design allows the acquisition of adequate statistics with an acceptable level of contamination from multiple scattering.  Figure~\ref{fig:mc} shows scattering distributions from a GEANT4~\cite{geant4} simulation of 10.8 keV recoils; the multiple scattering contributes less than $23\%$ of the total event rate between 5 and 16\,\kevr, and the position of the single scattering peak is not affected by the background.

\begin{figure*}[t!]
\includegraphics[width=0.92\columnwidth]{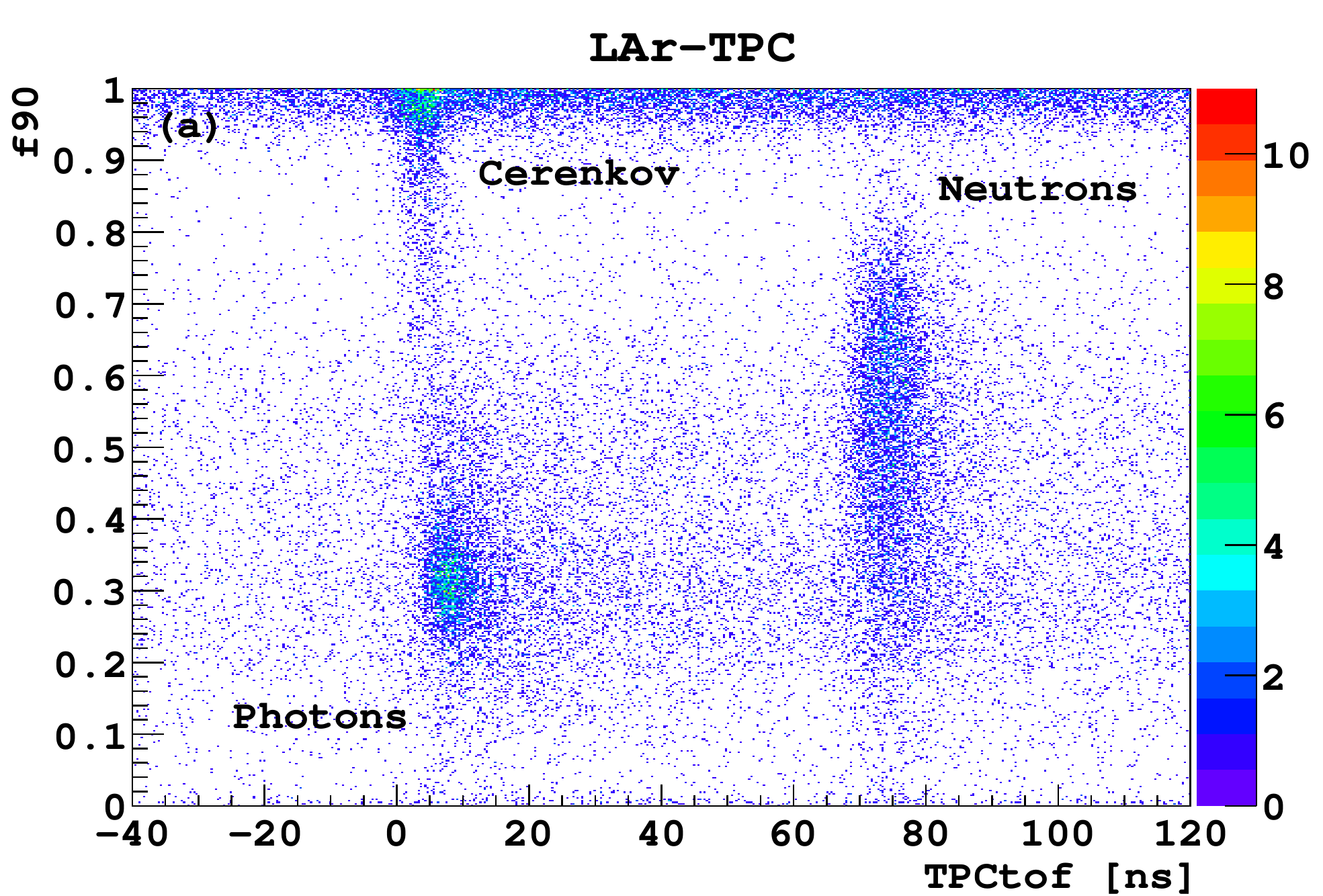}
\includegraphics[width=0.92\columnwidth]{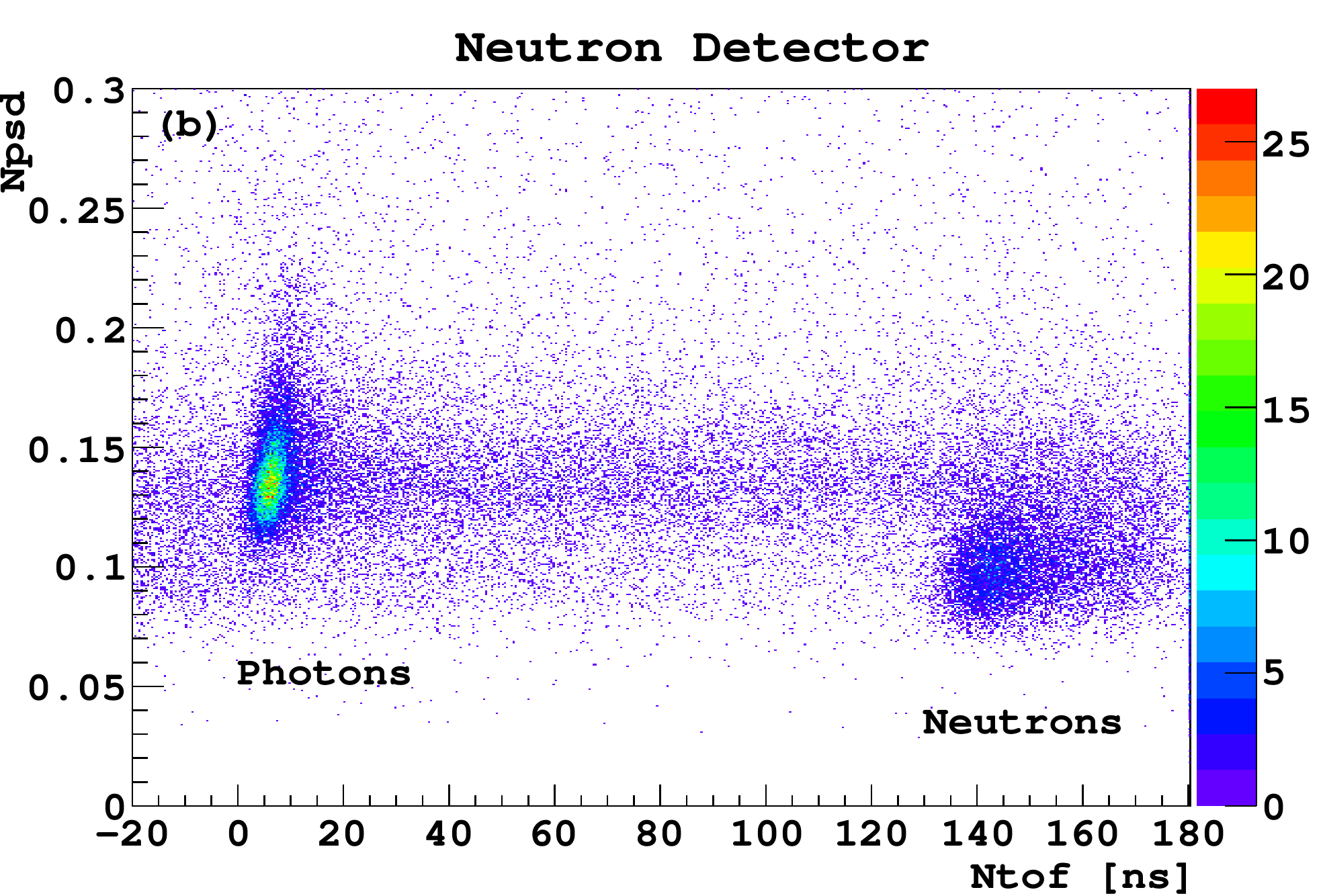}
\caption{\label{fig:tof-cuts}Distributions of pulse shape discrimination vs. time of flight for data taken in the 10.8\,\kevr\ configuration at 1000 V/cm. All variables are defined in the text. Panel (a) refers to the \lartpc\ and panel (b) to the neutron detectors.  }
\end{figure*}

\begin{figure*}[t!]
\includegraphics[width=0.95\columnwidth]{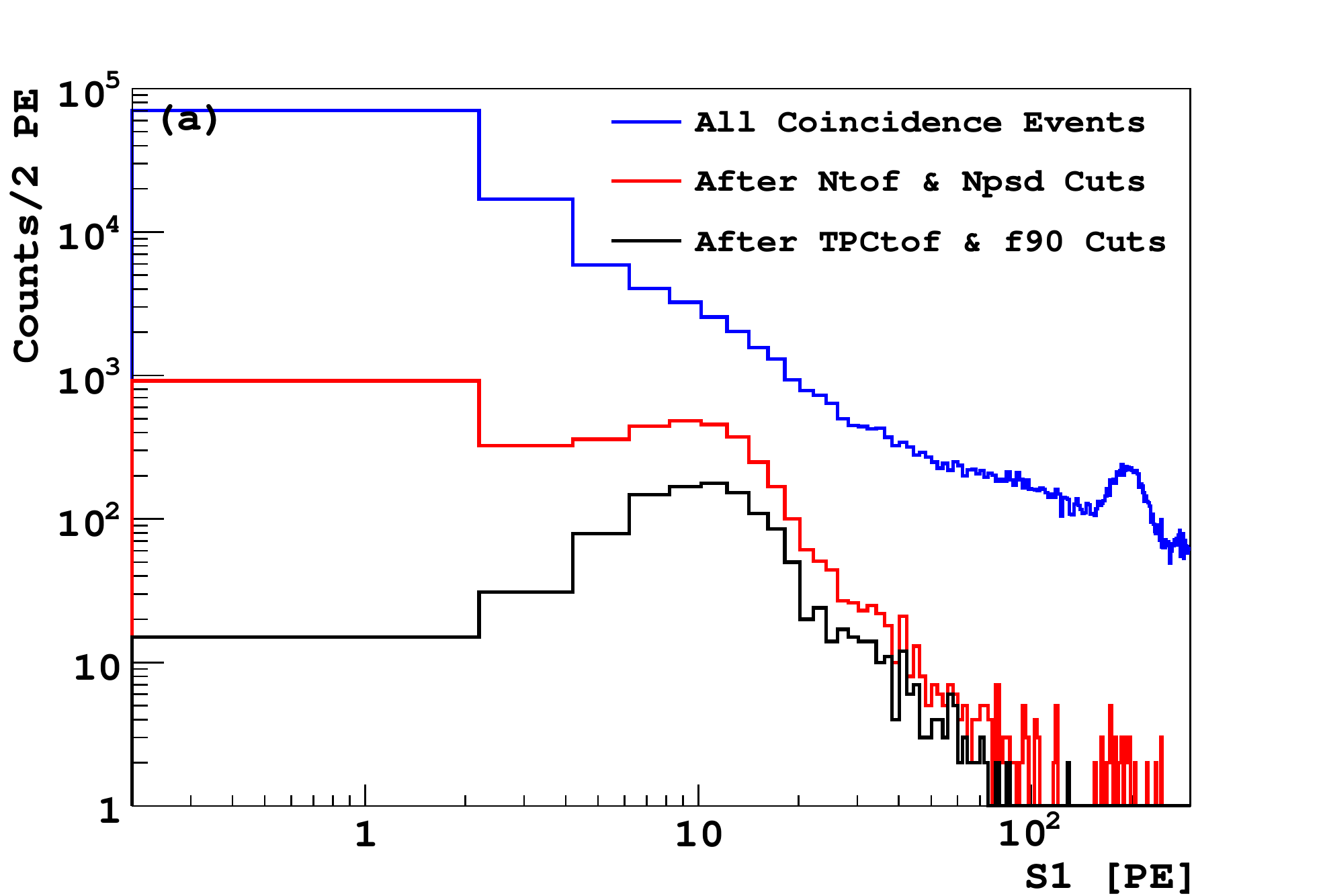}
\includegraphics[width=0.95\columnwidth]{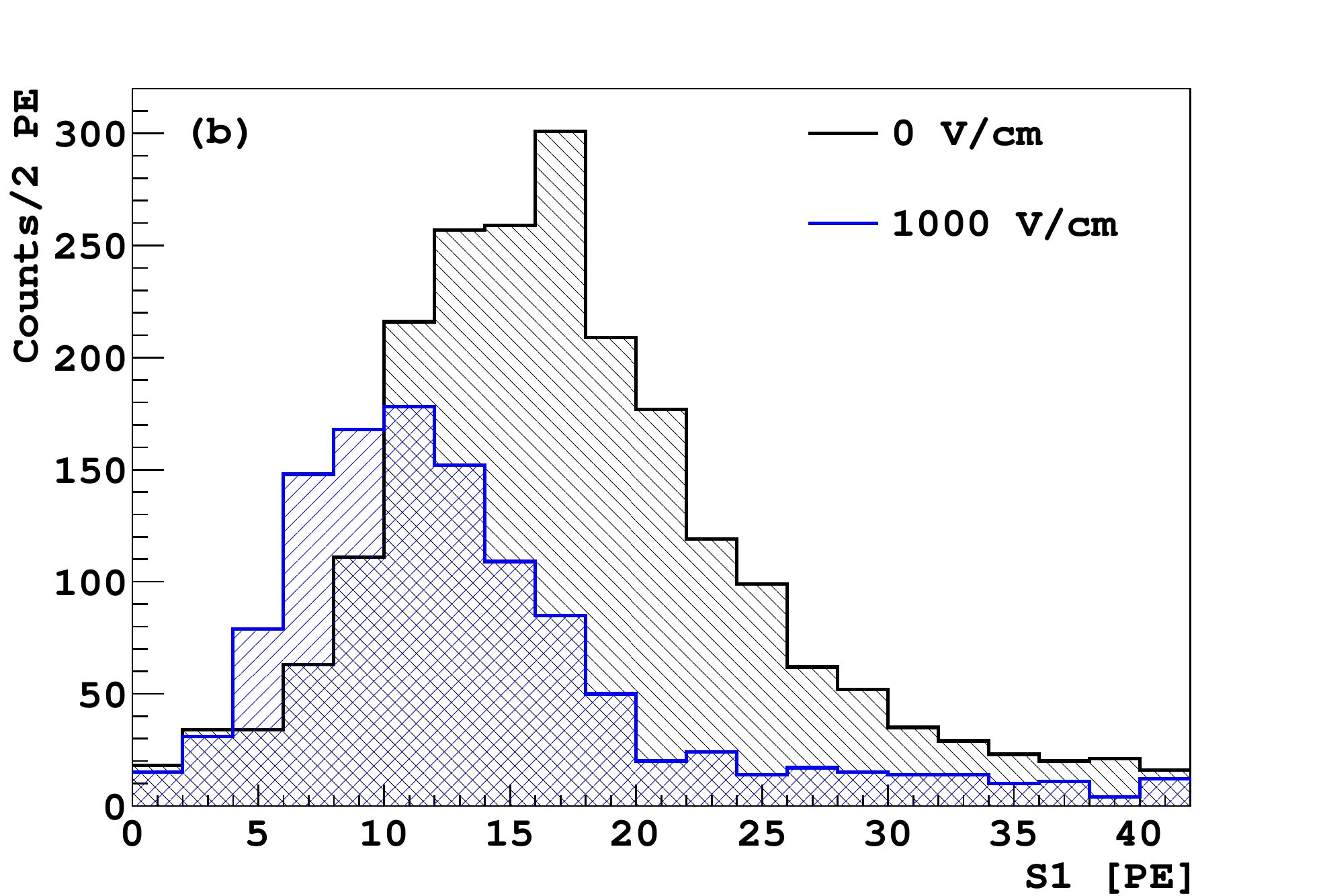}
\caption{\label{fig:s1}(a) Surviving primary scintillation light (S1) distributions for 10.8\,\kevr\ nuclear recoils as the neutron selection cuts described in the text are imposed sequentially.  Data were collected with a field of 1000 V/cm.  The high energy peak is from the \krthree\ source in use for continuous monitoring of the detector.  (b) Final S1 distributions for the 10.8\,\kevr\ nuclear recoils at null field at 1000 V/cm.}
\end{figure*}

\begin{figure*}[t]
\includegraphics[width=0.92\columnwidth]{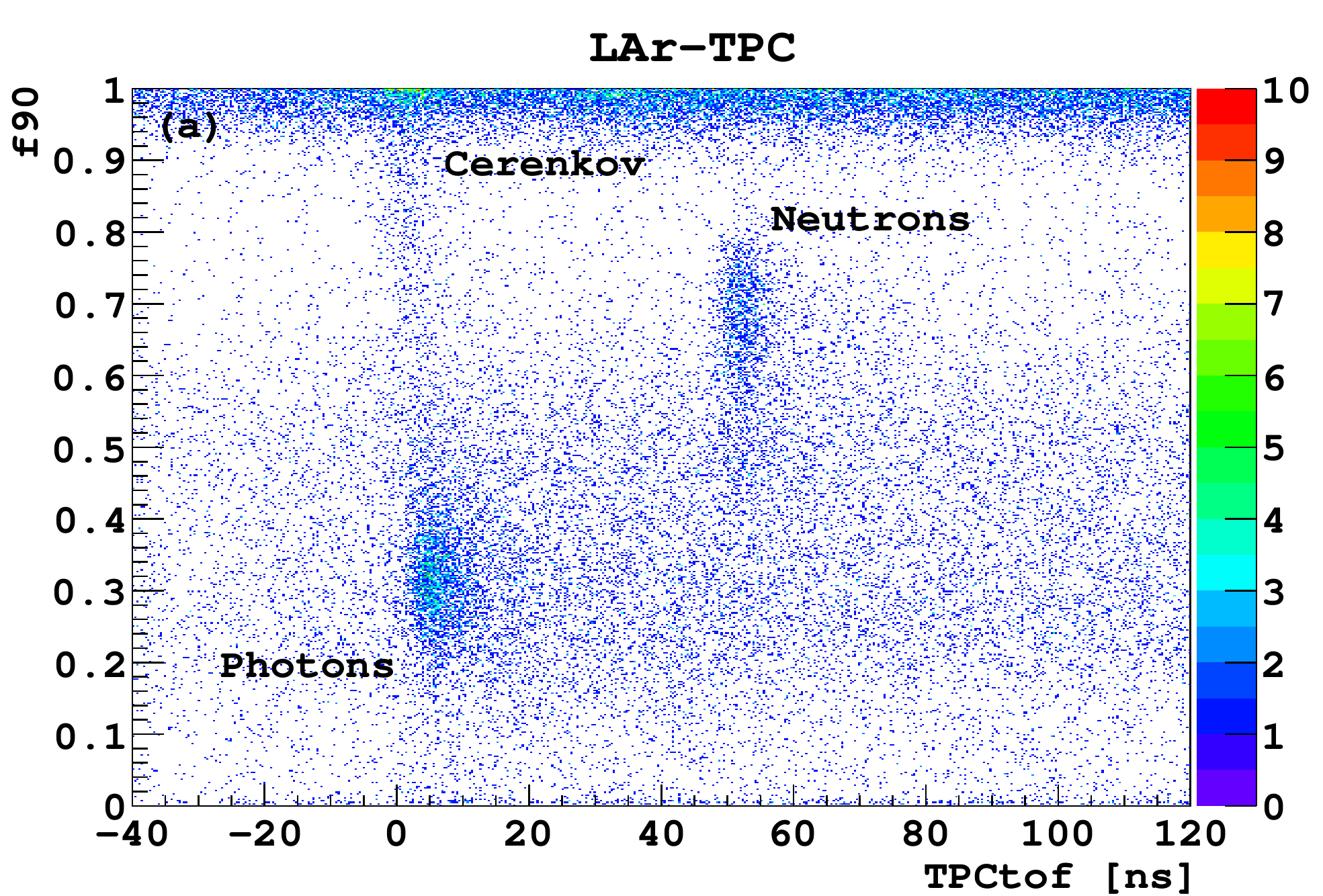}
\includegraphics[width=0.92\columnwidth]{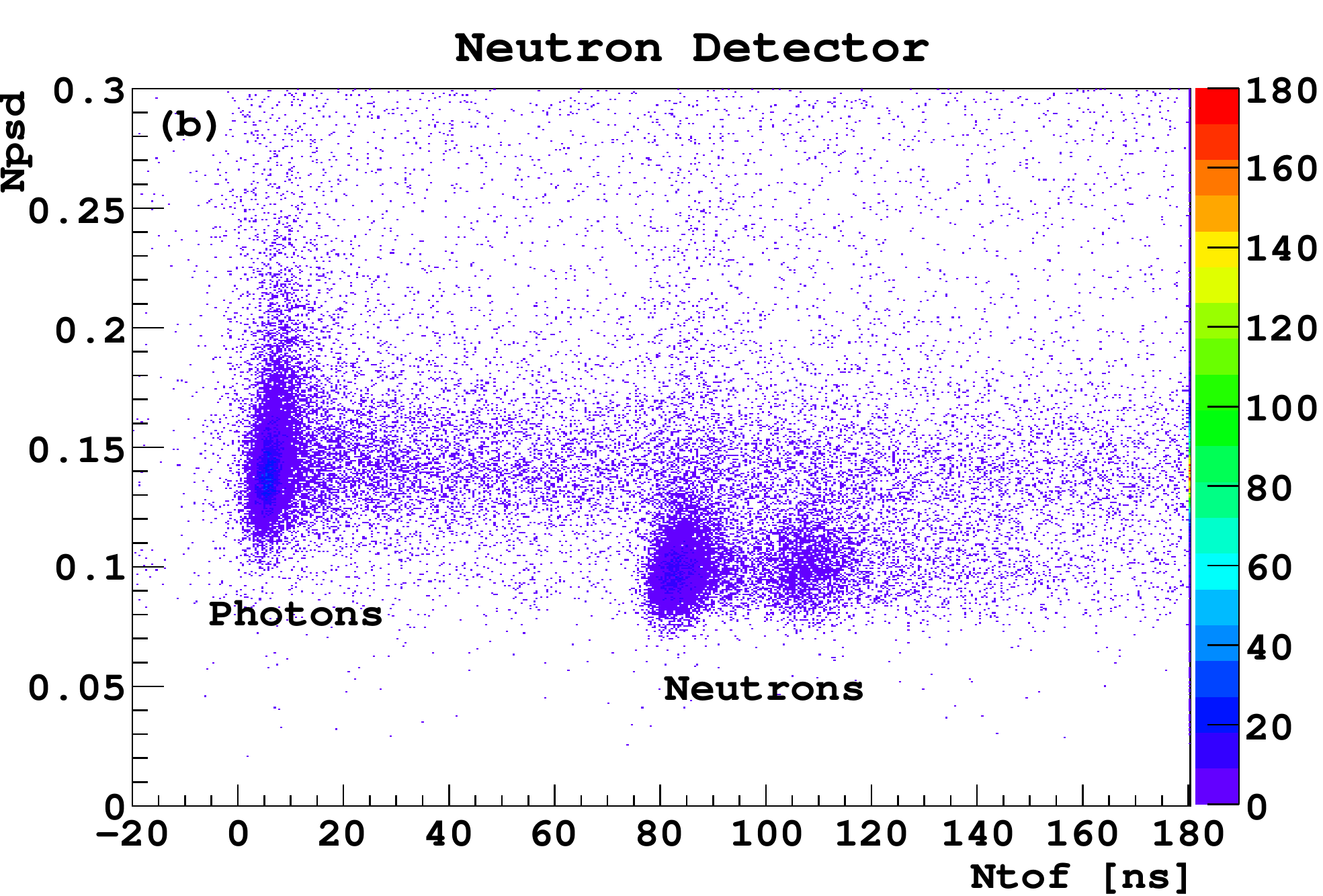}
\caption{\label{fig:tof-cuts_50keV}Distributions of pulse shape discrimination vs. time of flight for data taken in the 49.9\,\kevr\ configuration at 1000 V/cm. All variables are defined in the text. Panel (a) refers to the \lartpc\ and panel (b) to the neutron detectors. For these data, the polyethylene cylinders blocking the line-of-sight between the LiF target and the neutron detectors were removed due to geometric constraints. Accidental direct neutron coincidence events appear as the first "Neutron" blob around 80 ns in panel (b).  }
\end{figure*}

\begin{figure*}[t]
\includegraphics[width=0.95\columnwidth]{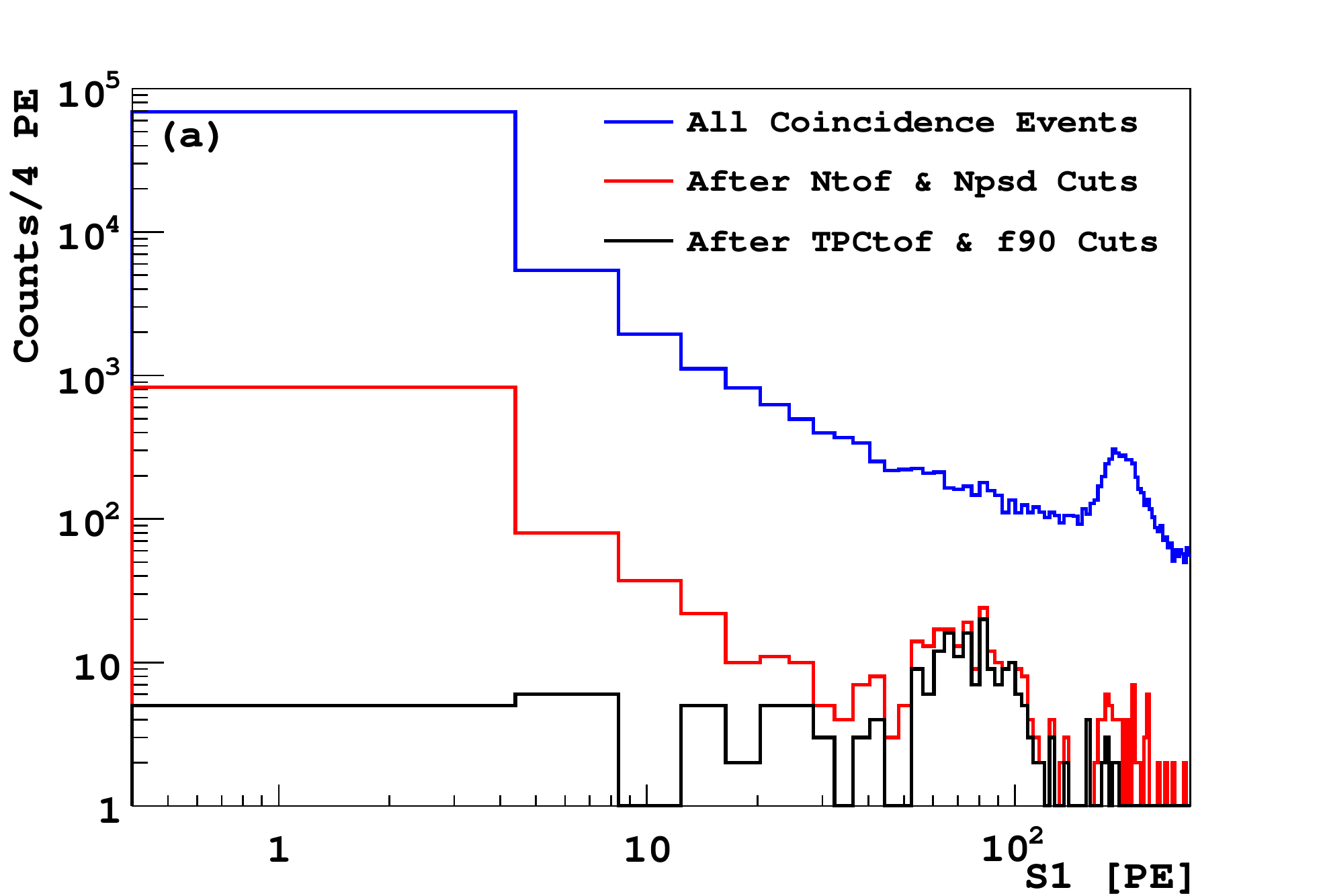}
\includegraphics[width=0.95\columnwidth]{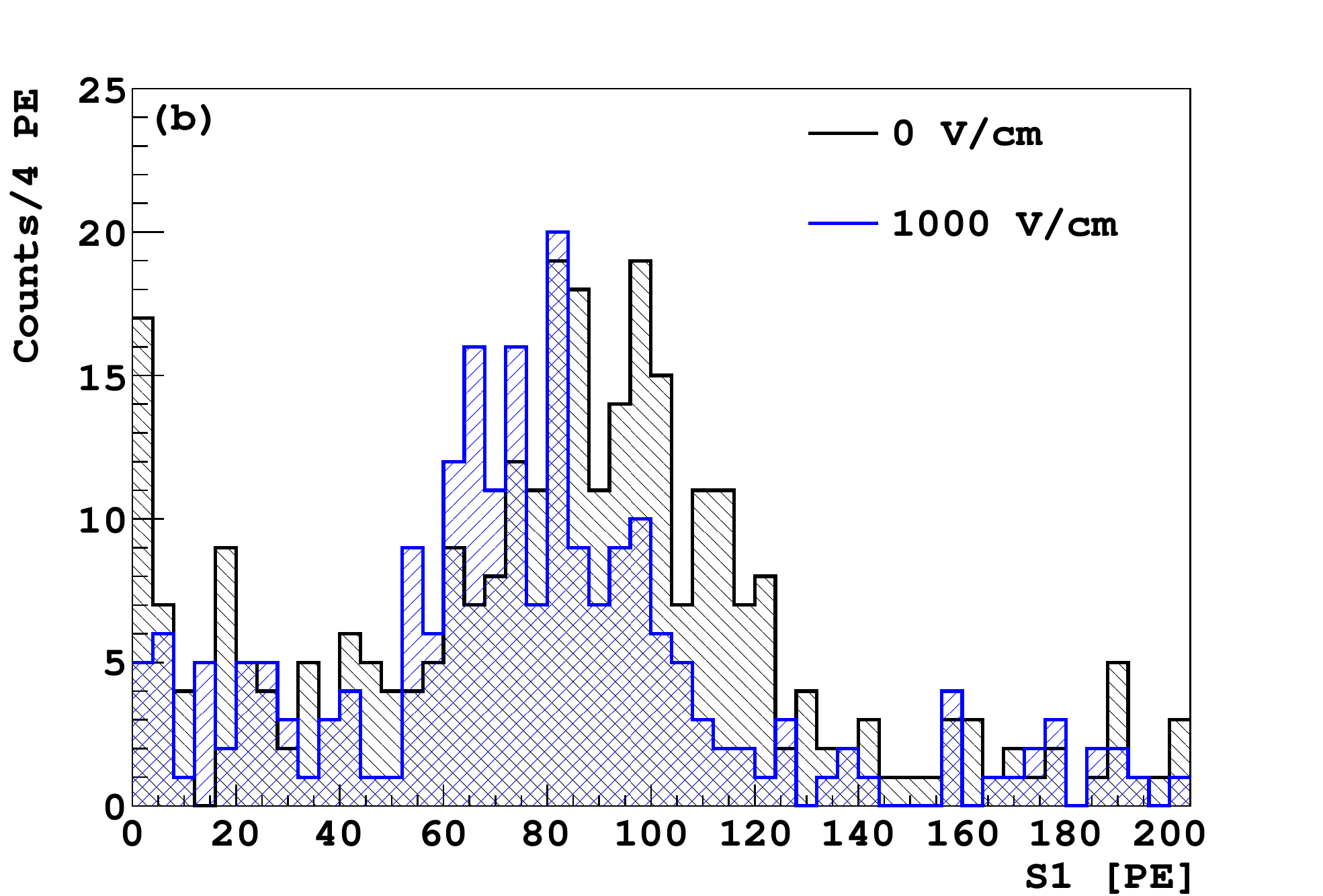}
\caption{\label{fig:s1_50keV}(a) Surviving S1 distributions for 49.9\,\kevr\ nuclear recoils as the neutron selection cuts described in the text are imposed sequentially.  Data were collected with a field of 1000 V/cm.  The high energy peak is from the \krthree\ source in use for continuous monitoring of the detector.  (b) Final S1 distributions for the 49.9\,\kevr\ nuclear recoils at null field at 1000 V/cm.}
\end{figure*}

To monitor the scintillation yield from the LAr, \krthree\ was continuously injected by including a \rbthree\,trap~\cite{kastens,lippincott_kr,manalaysay} in the recirculation loop.  \krthree\ has a half life of 1.82\,hours and decays via two sequential electromagnetic transitions with energies of 9.4 and 32.1\,keV and a mean separation of 222~\,ns. Because scintillation signals in LAr last for several microseconds~\cite{hitachi83}, we treat the two decays as a single event.  The activity of \krthree\ in the \lartpc\ was 1.2\,kBq.

The experiment trigger required a coincidence of either of the two \lartpc\ PMTs with either neutron detector.  In addition, we recorded events triggered by the \lartpc\ alone, consisting largely of \krthree\ events, at a rate of 12\,Hz.  The trigger thresholds of the \lartpc\ PMTs were set to $\sim$0.2\,photoelectrons (\pe).  The \lartpc\ trigger efficiency was measured to be above 90\% for pulses of 1\,\pe\ and greater using positron annihilation radiation from a \sodium\ source placed between the \lartpc\ and the neutron detector in a manner similar to that described in~\cite{plante}.

%The efficiency measurement was done using a	22 Na	source	(a	? +	emitter)	placed	between	the	LXe	detector and a sodium iodide NaI(Tl) detector. The back-to-back pair of 511 keV ? rays from the ?+ annihilation interact effectively at the same time in the LXe and the NaI(Tl) detectors. The NaI(Tl) detector was positioned such that the solid angle it subtended at the source was larger than the one subtended by the active LXe volume. This ensures that the whole active volume of the LXe detector is probed. For this measurement, the triggering signal consisted of the discriminated signal of the output of the NaI(Tl) detector. In addition to the signals of the LXe PMTs, the trigger signal of the normal LXe trigger was digitized with the flash ADC. The efficiency is inferred by computing the fraction of events accompanied by a LXe trigger signal as a function of their measured number of photoelectrons. Figure 5 shows the resulting trigger efficiency.

The stability of the complete light collection and conversion system is of critical importance to our measurements and was monitored in several ways.  The single \pe\ response of each PMT, determined using pulses in the tails of scintillation events, was continuously monitored and showed a slow decline of about 15$\%$ in the top PMT and $10\%$ in bottom PMT over the course of the 6 day run.  The stability of the entire system was assessed throughout the data taking by \krthree\ runs in the absence of drift field.  The position of the \krthree\ peak was measured to be 260\,\pe\ and varied by less than $\pm$2.4\% for the data presented here.  The short term stability during a run in presence of a drift field was checked with \krthree\ spectra accumulated every 15 minutes; these show negligible variations during a given run.

Our data include a collection of prompt events characterized by narrow pulses in the \lartpc\ PMTs with timing slightly earlier than photon-induced scintillation events (the LAr fast scintillation component has a decay time of $\sim$6\,ns~\cite{hitachi83}), and data taken with no liquid in the TPC have a similar collection of events. We take these signals to be Cerenkov radiation and therefore independent of scintillation processes in the argon and examine them to study any dependence of the apparatus response on the drift field. The spectrum of these events shows a peak at $\sim$80\,\pe\ which is stable within $\pm 2.5\%$  over all the electric field settings.

%We exploited prompt events from $\gamma$-ray interactions in the windows of the \lartpc\ and the PMTs (see Fig.~\ref{fig:tof-cuts}(a)) to study any dependence of the apparatus response on the drift field. These events are characterized by fast pulses in the \lartpc\ PMTs with timing slightly earlier than photon-induced scintillation events (the LAr fast scintillation component has a decay time of $\sim$6\,ns~\cite{lippincott}) and are therefore independent of scintillation in the argon.  This event spectrum shows a peak at $\sim$80\,\pe\ which is stable within $\pm 2.5\%$  over all the electric field settings.

The data acquisition system was based on 250\,MSPS waveform digitizers~\cite{caen}, which recorded signals from the \lartpc\ and neutron detectors and the accelerator RF signal.   The data were recorded using the "Daqman" data acquisition and analysis software~\cite{daqman}. The digitizer records were 16\,$\mu$s long including 5\,$\mu$s before the hardware trigger (used to establish the baseline).

Figure~\ref{fig:tof-cuts}(a) shows a scatterplot of the discrimination parameter \fno\ ~\cite{lippincott}, defined as the fraction of light detected in the first 90 ns of an event, vs. the time difference between the proton-beam-on-target and the TPC signal (\tpctof) for all events taken in the 10.8\,\kevr\ configuration at 1000 V/cm.  Beam-associated events with $\gamma$-like and neutron-like \fno\ are clustered near 10 and 75\,ns respectively, as expected given a speed of approximately 0.1$c$ for 604\,\kevr\ neutrons.  Cerenkov events are characterized by \fno\ close to~1.0 and $\gamma$-like timing.  The \krthree\ events appear with $\gamma$-like \fno\ uniformly distributed in \tpctof.  For the same events, Fig.~\ref{fig:tof-cuts}(b) shows a scatterplot of the neutron pulse shape discriminant (\npsd), defined as peak over area in the neutron detectors, vs. the time difference between  the proton-beam-on-target and the neutron detector signal (\ntof).  Neutron events scattering in the LAr-TPC cluster near a \npsd\ of 0.09 and a \ntof\ of 140\,ns, while \bg\ events cluster near a \npsd\ of 0.13 and a \ntof\ of 5\,ns.  Random coincidences from environmental backgrounds are visible at intermediate times.

We exploit the pulse-shape discrimination and timing information available to define a series of cuts. Figure~\ref{fig:s1}(a) shows the primary scintillation, or S1, spectra as these cuts are imposed in sequence.  The final spectrum at each drift field  and recoil energy is fit to a Gaussian and the mean value in the fit represents our measurement of the yield. Fig.~\ref{fig:s1}(b) shows final spectra at the 10.8\,\kevr\, configuration for 0 V/cm and 1000 V/cm.  For comparison, Figs.~\ref{fig:tof-cuts_50keV} and~\ref{fig:s1_50keV} show the same information for the 49.9\,\kevr\ data point. In this configuration, the polyethylene cylinders blocking the direct line of sight between the target and the neutron detectors were absent, resulting in a population of direct neutron events visible near 80 ns in Fig.~\ref{fig:tof-cuts_50keV}(b). Because the neutron beam energy was higher to produce these data (see Table~\ref{table:neutrons}), good scattering events cluster at shorter times of flight relative to the 10.8\,\kevr\ point (53 and 110\,ns in Figs.~\ref{fig:tof-cuts_50keV}(a) and (b) versus 75 and 140\,ns in Figs.~\ref{fig:tof-cuts}(a) and (b)).

\begin{figure}[t!]
\includegraphics[width=\columnwidth]{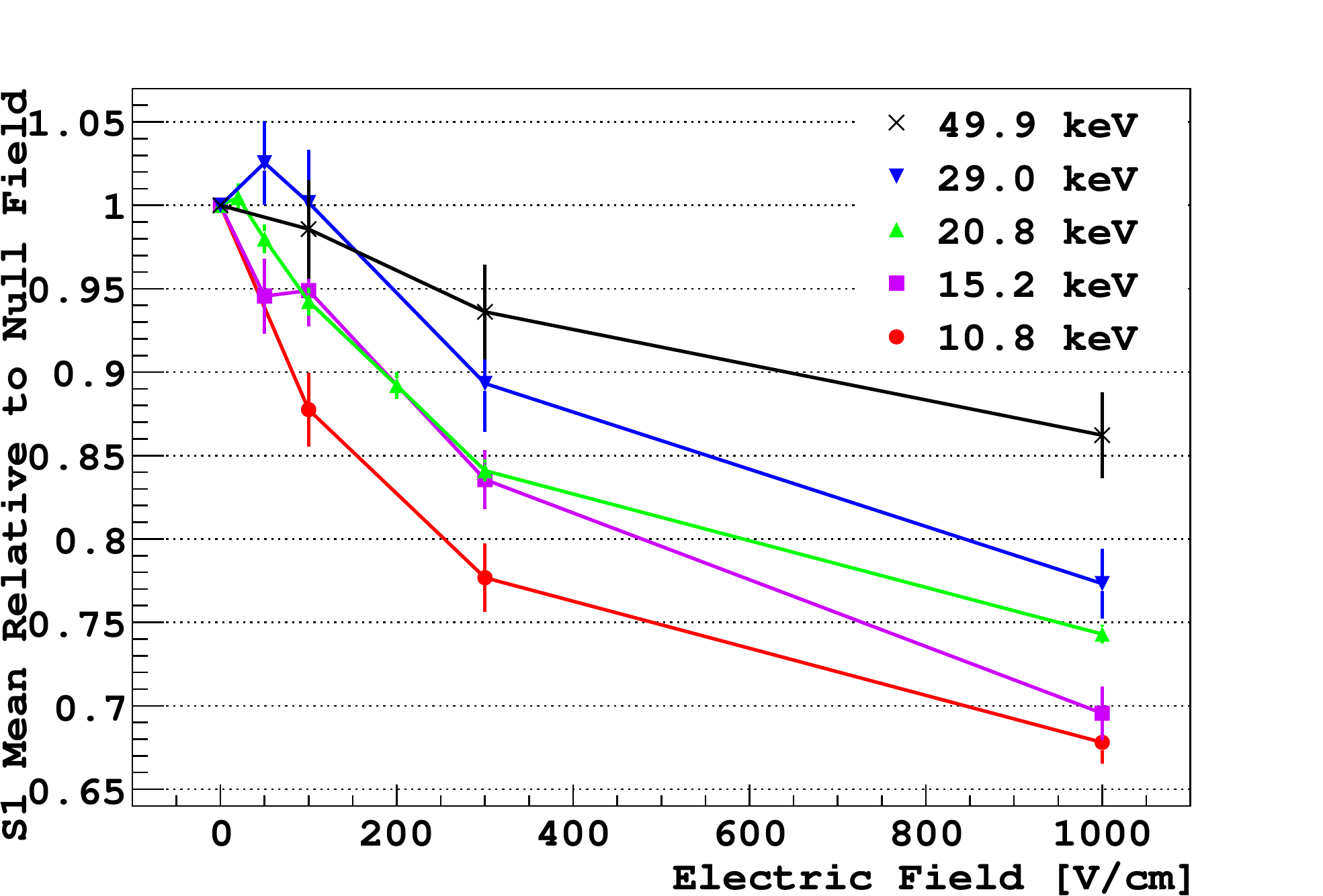}
\caption{\label{fig:ly}Variation of the S1 scintillation yield for 10.8 to 49.9\,\kevr\ nuclear recoils as a function of drift field normalized to the value at null field. Error bars on the data points taken at non-zero field are statistical only and include the uncertainty on the zero-field value.}
\end{figure}

The light yield normalized to the value at null field is shown as a function of drift field in Fig.~\ref{fig:ly} for five different recoil energies.  As can be seen, the signal yield decreases significantly as a function of the applied electric field, and this effect is more pronounced for lower energy recoils. At a field of 1\,kV/cm, the light yield is reduced by 32\% (14\%) for 10.8 (49.9)\,\kevr\ recoils. Such a significant dependence on field has not been previously reported. 

The presence of an external electric field can reduce the light yield by recoils in noble liquids largely because the external field reduces the probability of ion-electron recombination and subsequent de-excitation, one of the processes that ultimately produces light. The size of the effect can be expected to depend on the relative strengths of the external field and the field due to the ionization of the medium, the latter being determined by the total ionization energy loss and the density of the ionization along the path of the recoil. In general, low density tracks are more likely to feel the influence of an external electric field. For electrons, the reduction of light-yield by an external field in both argon and neon is well-known~\cite{shutt, Kubota1978}.  For our energies of interest ($<50$\,keV), we note that the argon nuclei are on the decreasing side of the Bragg peak (see Fig.~4 in~\cite{regenfus} for example), and therefore lower energy nuclear recoils have lower dE/dx. The simplest interpretation of our data is that a modest  external electric field reduces the recombination probability, and this reduction increases as the argon recoil energy decreases.

  To date, estimates of the sensitivity of LAr-TPC dark matter searches are based on the assumption that electric field has only a small, energy-independent effect on the light yield from nuclear recoils. Our result has important implications for the sensitivity estimates of LAr-TPC dark matter experiments and on the choice of operating parameters of those experiments.  
 While our results do not necessarily transfer to xenon, the quality of the data enabled by this technique may prompt similar investigations in liquid xenon. 
 
 %hile a great deal more data exist for xenon 
%Published dark matter limits from liquid xenon TPC searches are based on the same assumption.    and may prompt similar investigations of such effects in other noble liquids. 

%To summarize, we present a new technique for measuring the scintillation yield for nuclear recoils in LAr, using a dedicated \lartpc\ exposed to a pulsed, narrowband, neutron beam.  We have measured the light yield from nuclear recoils in LAr over a range of energies between 11 and 50\,\kevr\ and find an markeded dependence of the yield on drift field that increases at lower energies.
  %To date, estimates of the sensitivity of LAr-TPC dark matter searches are based on the assumption that electric field has only a small, energy-independent effect on the light yield from nuclear recoils. Published dark matter limits from liquid xenon TPC searches are based on the same assumption. Our result therefore has implications for the operating parameters of LAr-TPC dark matter searches and may prompt similar investigations of such effects in other noble liquids. 

\begin{acknowledgments}
We particularly thank the technical staff at Fermilab, and A.~Nelson of Princeton University and E.~Kaczanowicz of Temple University for their contributions to the construction of the SCENE apparatus.  We thank Dr.~G. Korga and Dr.~A. Razeto for providing the low-noise amplifiers used on the \lartpc\ PMT signals.  We thank Prof.~D.~N.~McKinsey, Dr.~S. Cahn, and K.~Charbonneau of Yale University for the preparation of the \krthree\ source.  Finally, we thank the staff at the Institute for Structure \& Nuclear Physics and the operators of the Tandem accelerator of the University of Notre Dame for their hospitality and for the smooth operation of the beam. 

The SCENE program is supported by NSF (U.S., Grants PHY-1314507, PHY-1242625, PHY-1211308, and associated collaborative Grants), DOE (U.S., Contract No. DE-AC02-07CH11359), and by the Istituto Nazionale di Fisica Nucleare (Italy ASPERA 1st common call, Darwin project).
\end{acknowledgments}

%%%%%%%%%

\end{document}